\documentclass[aps,twocolumn,nofootinbib]{revtex4}

\usepackage{graphics}
\usepackage{bm}
\usepackage{amssymb}

\begin{document}

\title{Orbital angular momentum spectrum of Bessel-Gaussian modes, \\ as generated by spontaneous parametric down-conversion}
\author{Filippus S. Roux}

\affiliation{CSIR National Laser Centre, P.O. Box 395, Pretoria 0001, South Africa}
\email{fsroux@csir.co.za}

\begin{abstract}
The Bessel-Gaussian modal spectrum, generated in spontaneous parametric down-conversion of a Gaussian pump beam, is considered. This is done by first deriving a general expression for the true probability of detecting specific transverse spatial modes for the pairs of photons generated in a spontaneous parametric down-conversion process. These expressions are applied for Bessel-Gaussian modes in type I phase matching with collinear, degenerate down-converted beams. The result shows that a broad orbital angular momentum spectrum can be obtained for Bessel-Gaussian modes under certain conditions.
\end{abstract}

\maketitle

\section{Introduction}

Although the connection between the spatial modal profile of an optical beam and orbital angular momentum (OAM) was initially made with specific reference to Laguerre-Gaussian beams \cite{allen}, the same property applies to any optical beam with a rotationally symmetric intensity profile. As a result, Bessel beams \cite{durnin1, durnin2} (or Bessel-Gaussian beams \cite{gori}) also have quantized amounts of OAM associated with them. Each photon in such a beam carries an amount of OAM equal to $\ell\hbar$, where $\ell$ is the azimuthal index of the mode.

The use of the OAM eigenstates of photons in quantum information science became attractive after it was shown, theoretically \cite{barbosa,franke} and experimentally \cite{zeil1}, that OAM is conserved during spontaneous parametric down-conversion (SPDC). A consequence of the conservation of momentum in the SPDC process, the conservation of OAM implies that the azimuthal indices of a pair of down-converted photons add up to that of the pump beam. As a result the pair of down-converted photons are naturally entangled in terms of their azimuthal indices or OAM. (This is not surprising, because entanglement in any spatial modal basis implies entanglement in every other spatial modal basis that are related to the former via unitary transformation. Since momentum conservation imply entanglement in terms of the plane wave basis, the entanglement in terms of OAM eigenstates is inevitable.) Presently, SPDC is the preferred method to prepare entangled photonic states for quantum information processing.

Entanglement is a desirable property in quantum information application. It is used in quantum ghost imaging \cite{pittman1}, quantum cryptography \cite{qkd0,qkd1} and in quantum computing algorithms \cite{nc}. Pairs of photons can also be entangled in terms of polarization, using SPDC with type II phase matching. In this case the Hilbert space is two-dimensional and the photonic quantum states (qubits) are all represented on a Bloch sphere. On the other hand, any transverse spatial modal basis defines an infinite dimensional Hilbert space, similar to the plane wave basis, which it replaces. A large Hilbert space can represent more information in quantum information applications.

Due to the increased information capacity and the fact that SPDC naturally produces OAM entanglement, the use of OAM eigenstates for quantum information processing became an attractive proposition. Currently the Laguerre-Gaussian modes are the popular choice for theoretical analyses of quantum information systems based on OAM. Ironically, the modal basis that are actually used in quantum information experiments are very seldom true Laguerre-Gaussian modes. Although, the helical phase of these modes can be manipulated and detected using linear optical systems, the manipulation of the full Laguerre-Gaussian modes is not without its challenges. The processing and detection of these modes are made complicated by the fact that they carry another index --- the radial index --- that governs the radial shape of the intensity profile. Each azimuthal index is therefore not associated with a unique quantum state, but with an infinite subspace of the total Hilbert space. If the radial part of its mode changes, a photon propagating through an optical system may suffer a loss of the quantum entanglement, even if the azimuthal index is unaffected. 

\begin{figure}[ht]
\centerline{\scalebox{1}{\includegraphics{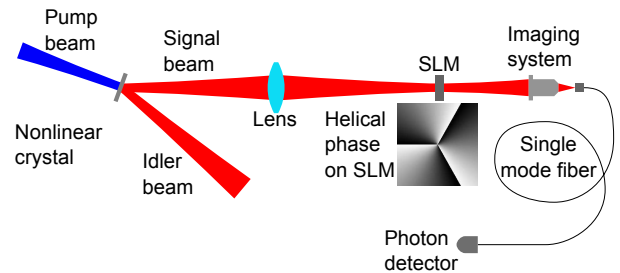}}}
\caption{Experimental setup to detect OAM eigenstate after SPDC, showing one beam with a helical phase of $\ell=3$.}
\label{demodoam}
\end{figure}

A widely used experimental technique for the detection of OAM eigenstates is to use a spatial light modulator (SLM) that contains the helical phase associated with a specific azimuthal index \cite{yao} (See Fig.~\ref{demodoam}). With this SLM the beam is (de)modulated with the complex conjugate of its helical phase function, thus removing the helical phase from the beam. The resulting beam is then coupled into a single mode fibre to extract the pure Gaussian modal component. The light inside the fibre is guided onto a photon detector to register the probability for the observed quantum state in terms of a photon count.

If the mode prior to the SLM is a pure Laguerre-Gaussian mode with a nonzero azimuthal index the overlap with the mode of the single mode fibre, which resembles a pure Gaussian, is very small, leading to a very small coupling efficiency and very small photon count. The reason for the low coupling efficiency is that the radial profile of a Laguerre-Gaussian mode with a nonzero azimuthal index does not resemble a pure Gaussian. For this reason most quantum information experiments do not employ actual Laguerre-Gaussian modes. Since what is coupled into the single mode fibre is very close to a pure Gaussian, the mode just behind the SLM that contributes to the observed photon count must also resemble a pure Gaussian.

Effectively the physical modes that are being used can be viewed as a projected subspace of the full Hilbert space of Laguerre-Gaussian modes. The azimuthal index is retained, but the radial dependence is projected onto something that resembles a pure Gaussian. This projection operation is the combined effect of the optical system and the detection process that is used in the quantum experiment. Imperfections of the optical system, such as its finite transfer function, together with the effect of the phase singularity of the helical phase on the intensity profile, produce uncontrolled and often unpredictable effects on the radial part of the mode. The radial part of the modal content of the quantum states is therefore often poorly defined and the effect thereof on the quantum state during quantum processing is often simply ignored. The nature of the radial part of the mode does not only affect the coupling efficiency and therefore the photon count, but also, more importantly, the fidelity of the quantum state. 

\begin{figure}[ht]
\centerline{\scalebox{1}{\includegraphics{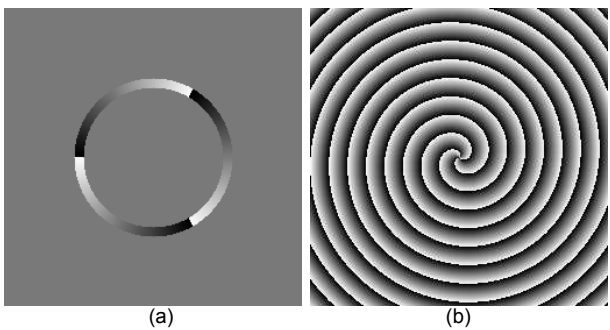}}}
\caption{Functions for the generation of Bessel-Gaussian modes. (a) Slit-ring aperture with a helical phase of $\ell=3$. (b) Conical phase function with a helical phase of  $\ell=3$.}
\label{bgmodgen}
\end{figure}

As the complexity of quantum information systems increase, the requirements for the fidelity of the quantum states can be expected to become more demanding. The radial dependence of the spatial modes will have to be treated with more care. To this end, Bessel-Guassian (BG) modes \cite{gori} provide better control of the radial dependence. The radial index associated with Laguerre-Gaussian modes is replaced by a continuous scaling parameter for the radial part of the BG modes. In practice the radial part of a BG mode is generated in the Fourier domain by an annulus or a ring-slit aperture, as shown in Fig.~\ref{bgmodgen}(a), the radius of which determines the scaling parameter. Alternatively, BG modes can be generated by a conical phase function, as shown in Fig.~\ref{bgmodgen}(b), where the implied conical angle determines the scaling parameter. Both these methods are combined with a helical phase function to determine the azimuthal index of the BG mode.

The conical phase method [Fig.~\ref{bgmodgen}(b)] is more efficient because in this case both the azimuthal and radial dependences of the mode are determined by phase functions, which do not introduce a loss in optical power. In contrast, to define the radial dependence of Laguerre-Gaussian modes one would need to perform complex amplitude modulation \cite{cam}, which results in the loss of optical power.

The relative easy and the accuracy with which one can manipulate the azimuthal and radial dependences of BG modes make them a favorable choice for high-fidelity quantum information systems. The question is whether one can prepare suitable entangled states in terms of the BG basis. We already know that SPDC generates states that are entangled in terms of their azimuthal indices, but how does the choice of the radial dependence of the modal basis affect the coefficients in the OAM expansion of these states? It is preferable that these coefficients have the same magnitude, because that represents a quantum state that is maximally entangled. As a result one prefers a broad flat spectrum in terms of the azimuthal index (i.e.\ a large spiral bandwidth \cite{torres1}).

In this paper we investigate the spectrum of BG modes generated by the SPDC process. Previously, the spiral bandwidth was calculated for the Laguerre-Gaussian modes \cite{torres1,miatto,ling}, but these calculations mostly focused only on the shape of the spectrum, discarding the magnitude. Ling {\it et al.} \cite{ling} attempted an analysis that gives the magnitude of the process, however their analysis made several assumptions and simplification that eventually lead to expressions that are of limited use. We develop the formalism for this calculation from first principles (Sec.~\ref{trapro}) and then simplify the expression (Sec.~\ref{simple}) by considering certain special cases, making all assumptions explicit. In the end we obtain expressions that not only provide the shape of the spectrum, but also its magnitude. Using a generating function for the angular spectra of the BG modes (Appen.~\ref{besgauss}), we then apply the expressions in the case where the modal basis is the BG modes (Sec.~\ref{bgspek}) and show that one can obtain a suitably flat spectrum for particular choices of the parameters. We end with some conclusions in Sec.~\ref{concl}.

\section{\label{trapro}Transition probability}

The OAM spectrum after SPDC is determined by the probability to measure a pair of down-converted photons in a particular final state (pair of transverse spatial modes). If we assume that the quantum state after the crystal is given by a density operator $\rho$, then the probability for this density operator to contain a particular final state is given by the trace ${\rm tr} \{ \rho \rho_f \}$ over the product of this density matrix with some final pure state $\rho_f = |\Psi_f\rangle \langle\Psi_f|$ that we are looking for. Assuming that the density operator evolves according to
\begin{equation}
i\hbar\ \partial_t \rho(t) = \left[ \rho(t), H_I(t) \right] ,
\end{equation}
in the interaction picture, where $H_I(t)$ is the interaction Hamiltonian, and that the initial states is also pure $\rho(t_0) = | \Psi_{in} \rangle \langle \Psi_{in} |$, we find that the trace can be expressed by
\begin{equation}
{\rm tr}\{ \rho \rho_f \} = \left| {1\over\hbar} \int_{t_0}^t \langle \Psi_f | H_I(t_1) | \Psi_{in} \rangle\ {\rm d} t_1 \right|^2 .
\label{rhoint}
\end{equation}
The quantity to be considered is, therefore, the scattering amplitude ${\cal M}$, given by
\begin{equation}
{\cal M} = {1\over\hbar} \int_{t_0}^t \langle \Psi_f | H_I(t_1) | \Psi_{in} \rangle\ {\rm d} t_1 .
\label{mint}
\end{equation}
If we view the SPDC process as a Feynman-diagram, then  Eq.~(\ref{mint}) represents the vertex. Implicit in this interaction is the requirement to conserve momentum and energy. Although similar to Fermi's golden rule, which was used in Ref.~\cite{ling} to compute the OAM spectrum, the expression in Eq.~(\ref{mint}) also contains an integral over time and, when substituted into Eq.~(\ref{rhoint}), yields a true probability and not a transition rate.

\subsection{Decoupling the overlap integral}

The initial and final states are `on shell' in the sense that the momentum and energy of these states obey the vacuum dispersion relation (shell condition)  $\omega=c|{\bf k}|$. As a result one can insert identity operators, defined in terms of the momentum basis,
\begin{equation}
1 = \sum_s \int |{\bf k},s\rangle \langle{\bf k},s|\ {{\rm d}^3 k\over\omega\ (2\pi)^3} ,
\label{ident}
\end{equation}
where the momentum basis elements obey the orthogonality condition \cite{ps} 
\begin{equation}
\langle {\bf k}_1,r|{\bf k}_2,s\rangle = \omega\ (2\pi)^3 \delta_{r,s}\ \delta({\bf k}_1-{\bf k}_2) ,
\label{inprodk0}
\end{equation}
with $\delta(\cdot)$ being the Dirac delta function. The $\omega$-factor in Eq.~(\ref{inprodk0}) comes from the requirement for Lorentz invariance \cite{ps}.

Identity operators, as defined in Eq.~(\ref{ident}), are now inserted for the pump, signal and idler beams, respectively. The initial state then becomes
\begin{equation}
|\Psi_i\rangle = \sum_r \int |{\bf k}_3,r\rangle \Psi_p({\bf k}_3,r)\ {{\rm d}^3 k_3\over\omega_3\ (2\pi)^3} ,
\label{intoes}
\end{equation}
where
\begin{equation}
\Psi_p( {\bf k}_3,r)=\langle {\bf k}_3,r|\Psi_{in}\rangle ,
\label{indef}
\end{equation}
and the final state becomes
\begin{eqnarray}
\langle \Psi_f | & = & \sum_{p,q} \int \Psi_s^*({\bf k}_1,p) \Psi_i^*({\bf k}_2,q) \nonumber \\ 
& & \times \langle{\bf k}_1,p,{\bf k}_2,q|\ {{\rm d}^3 k_1\over\omega_1\ (2\pi)^3}\ {{\rm d}^3 k_2\over\omega_2\ (2\pi)^3} ,
\label{uitoes}
\end{eqnarray}
where
\begin{equation}
\Psi_s^*({\bf k}_1,p) \Psi_i^*({\bf k}_2,q) = \langle\Psi_f|{\bf k}_1,p,{\bf k}_2,q\rangle .
\label{findef}
\end{equation}
Substituting Eqs.~(\ref{intoes}) and (\ref{uitoes}) into Eq.~(\ref{mint}), we obtain
\begin{eqnarray}
{\cal M} & = & \sum_{p,q,r} \int \Psi_s^*({\bf k}_1,p) \Psi_i^*({\bf k}_2,q)\ {\cal V}\ \Psi_p({\bf k}_3,r) \nonumber \\ 
& & \times {{\rm d}^3 k_1\over\omega_1\ (2\pi)^3}\ {{\rm d}^3 k_2\over\omega_2\ (2\pi)^3}\ {{\rm d}^3 k_3\over\omega_3\ (2\pi)^3} ,
\label{mamp}
\end{eqnarray}
where the phase space integrations share the same integral sign and
\begin{equation}
{\cal V} = {1\over\hbar} \int_{t_0}^t \langle{\bf k}_1,p,{\bf k}_2,q |H_I(t_1)|{\bf k}_3,r\rangle\ {\rm d} t_1 ,
\label{pkwant}
\end{equation}
which is interpreted as the Feynman rule for the vertex.

In this way we have now decoupled the calculation associated with the vertex itself from the overlap of the particular spatial modes of the pump, signal and idler. One can now evaluate the integrals for the vertex rule once and for all and afterwards use the same result for any particular set of initial and final states given in terms of the transverse spatial modes of the pump, signal and idler beams.

\subsection{The interaction Hamiltonian}

Assuming that one can neglect the frequency dependence of the second order nonlinear susceptibility $\chi^{(2)}$, the interaction Hamiltonian is given by
\begin{equation}
H_I(t) = \int_V \epsilon_0\ \chi^{(2)}_{abc} E_a^{(p)} E_b^{(s)} E_c^{(i)}\ {\rm d}^3 x ,
\label{inth}
\end{equation}
where the integration runs over the volume $V$ of the nonlinear medium. The electric field vectors are given in terms of their components, which means that the indices represent the three spatial coordinates $a,b,c=\{x,y,z\}$ and repeated indices are summed over. The expression for the quantized electric field is given by
\begin{widetext}
\begin{eqnarray}
{\bf E}({\bf x},t) & = & {\bf E}^{(-)}({\bf x},t) + {\bf E}^{(+)}({\bf x},t) \nonumber \\
 & = & \sum_s \int \sqrt{n_s({\bf k}) \hbar\omega\over 2\epsilon_0}\ \eta_s(\hat{k})\ a_s({\bf k}) \exp[i\omega t - i n_s({\bf k}) ({\bf k}\cdot{\bf x})]\ {{\rm d}^3 k\over (2\pi)^3} \nonumber \\
& & + \sum_s \int \sqrt{n_s({\bf k}) \hbar\omega\over 2\epsilon_0}\ \eta_s^{\dag}(\hat{k})\ a_s^{\dag}({\bf k}) \exp[-i\omega t + i n_s({\bf k}) ({\bf k}\cdot{\bf x})]\ {{\rm d}^3 k\over (2\pi)^3} .
\label{eveld0}
\end{eqnarray}
\end{widetext}
where $\eta_s(\hat{k})$ represents the polarization vector for a given spin state, denoted by $s$, and a given propagation direction, denoted by $\hat{k}$. We made the spin state dependent refractive index $n_s({\bf k})$ of the medium explicit by expressing the electric field in terms of the vacuum momentum basis, which obeys the vacuum dispersion relation $\omega=c|{\bf k}|$. The creation and annihilation operators obey the following commutation relations
\begin{equation}
\left[ a_s({\bf k}_1), a_r^{\dag}({\bf k}_2) \right] = (2\pi)^3 \delta_{s,r}\ \delta( {\bf k}_1- {\bf k}_2 ) ,
\label{commut0}
\end{equation}
which implies that, for consistency with Eq.~(\ref{inprodk0}),
\begin{eqnarray}
\langle {\bf k},s| & = & \sqrt{\omega}\ \langle 0 | a_s({\bf k})  \nonumber \\
| {\bf k}, s \rangle & = & \sqrt{\omega}\ a_s^{\dag}({\bf k}) | 0 \rangle .
\label{toest}
\end{eqnarray}

\subsection{Simplifying the vertex rule}

When the electric field operates on one of the basis states defined in Eq.~(\ref{toest}), one obtains
\begin{widetext}
\begin{eqnarray}
{\bf E}({\bf x},t) |{\bf k}',r\rangle
& = & \left\{ \sum_s \int \sqrt{n_s({\bf k}) \hbar\omega\over 2\epsilon_0} \left[ \eta_s(\hat{k}) a_s({\bf k}) \exp(i\omega t - i n_s({\bf k}){\bf k}\cdot{\bf x}) + {\rm h.c.} \right] {{\rm d}^3 k\over (2\pi)^3} \right\} \sqrt{\omega'}\ a_r^{\dag}({\bf k}') |0\rangle \nonumber \\
& = & \sqrt{n_s({\bf k}') \hbar\over 2\epsilon_0}\ \eta_r(\hat{k}') \exp(i\omega' t - i n_s({\bf k}') {\bf k}'\cdot{\bf x})\ \omega'\ |0\rangle + ... ,
\label{emom}
\end{eqnarray}
\end{widetext}
where the additional terms that we neglect are two photon states that are orthogonal to the final state. Substituting Eqs.~(\ref{inth}), (\ref{eveld0}) and (\ref{toest}) into Eq.~(\ref{pkwant}) and applying Eq.~(\ref{emom}), one obtains
\begin{eqnarray}
{\cal V} & = & 6 \sqrt{\hbar\over 8\epsilon_0} \int_{t_0}^t \int_V \chi^{(2)}_{abc} \eta_p^{a*}(\hat{k}_1) \eta_q^{b*}(\hat{k}_2) \eta_r^c(\hat{k}_3) \nonumber \\ 
& & \times \sqrt{n_p({\bf k}_1) n_q({\bf k}_2) n_r({\bf k}_3)}\ \omega_1 \omega_2 \omega_3 \nonumber \\ 
& & \times \exp(i \Delta \omega t_1 - i \Delta {\bf k}\cdot{\bf x})\ {\rm d}^3 x\ {\rm d} t_1
\label{vdef}
\end{eqnarray}
where the 6 comes from all possible ways in which the electric field operators in the interaction Hamiltonian can be contracted with the initial and final states, together with the fact that the second order nonlinear susceptibility is symmetric with respect to any permutation of the indices; ${\bf k}_1$, ${\bf k}_2$ and ${\bf k}_3$ denote the propagation vectors associated with the signal, idler and pump bases, respectively; $p$, $q$ and $r$ denote the spin states of the signal, idler and pump beams, respectively; $a$, $b$ and $c$ are spatial indices that are summed over as in Eq.~(\ref{inth}); 
\begin{equation}
\Delta \omega = \omega_3-\omega_1-\omega_2
\label{dwdef}
\end{equation}
and
\begin{equation}
\Delta {\bf k} = n_r({\bf k}_3){\bf k}_3-n_p({\bf k}_1){\bf k}_1-n_q({\bf k}_2){\bf k}_2 .
\label{dkdef}
\end{equation}

Only the exponential function in Eq.~(\ref{vdef}) contains dependencies on the spatial coordinates and time. The volume $V$ is assumed to be infinite along the transverse dimensions $x$ and $y$, and finite along the general propagation direction $z$. Therefore
\begin{eqnarray}
\int_V \exp( -i \Delta {\bf k}\cdot{\bf x})\ {\rm d}^3 x & = & (2\pi)^2 L\ \delta(\Delta k_x) \delta(\Delta k_y) \nonumber \\ 
& & \times S(\Delta k_z L) ,
\end{eqnarray}
where
\begin{equation}
S(\Delta k_z L) = {2\over\Delta k_z L} \sin \left( {\Delta k_z L\over 2}\right) = {\rm sinc} \left( {\Delta k_z L\over 2\pi}\right) 
\label{sfunc}
\end{equation}
and $L$ is the thickness of the nonlinear medium along the $z$ direction.

If there are no restrictions on the time for the interaction, other than that which is imposed by the spatial extent of the medium, we can extend the time interval to infinity, which gives
\begin{equation}
\int_{-\infty}^{\infty} \exp(i \Delta \omega t_1)\ {\rm d} t_1 = 2\pi \delta(\Delta \omega) .
\end{equation}
(Other restrictions could include the short duration of the laser pulse. Here we assume a CW source.)

The expression for the vertex rule then becomes  
\begin{eqnarray}
{\cal V} & = & c^{3/2} L\ g_{pqr}({\bf k}_1,{\bf k}_2,{\bf k}_3) {\omega_1 \omega_2 \over \omega_3} (2\pi)^3 \nonumber \\ 
& & \times \delta(\Delta \omega) \delta(\Delta k_x) \delta(\Delta k_y) S(\Delta k_z L) ,
\label{feynman}
\end{eqnarray}
where $g_{pqr}$ is a dimensionless effective nonlinear coefficient for the vertex, given by 
\begin{eqnarray}
g_{pqr}({\bf k}_1,{\bf k}_2,{\bf k}_3) & = & \chi^{(2)}_{abc} \eta_p^{a*}(\hat{k}_1) \eta_q^{b*}(\hat{k}_2) \eta_r^c(\hat{k}_3) \sqrt{\hbar\over 2c^3\epsilon_0} \nonumber \\ 
& & \times 3 \omega_3^2 \sqrt{n_p({\bf k}_1) n_q({\bf k}_2) n_r({\bf k}_3)}
\end{eqnarray}
and which depends on the polarization states of the signal, idler and pump, as denoted by the subscripts $p$, $q$ and $r$, respectively. Henceforth, we'll distinguish between the different combinations of spin states that give phase matching, leading to either type I or type II phase matching. The expressions for the nonlinear coefficient $g$ and $\Delta {\bf k}$ in these two cases are provided in Appen.~\ref{pmc} for the benefit of the reader.

The vertex rule, as expressed in Eq.~(\ref{feynman}), contains no more integrals. It does contain a number of Dirac delta functions that impose energy and momentum conservation (to some extent) and a sinc-function $S(\Delta k_z L)$ that imposes the phase matching condition.

\subsection{General expression}

We now substitute ${\cal V}$, given in Eq.~(\ref{feynman}), into ${\cal M}$, given in Eq.~(\ref{mamp}), to obtain
\begin{eqnarray}
{\cal M} & = & c^{3/2} L \int g({\bf k}_1,{\bf k}_2,{\bf k}_3) \Psi_s^*({\bf k}_1) \Psi_i^*({\bf k}_2) \Psi_p( {\bf k}_3) \nonumber \\ & & \times (2\pi)^3 {\delta(\Delta \omega) \over\omega_3^2} \delta(\Delta k_x) \delta(\Delta k_y) S(\Delta k_z L) \nonumber \\ & & \times {{\rm d}^3 k_1\over (2\pi)^3}\ {{\rm d}^3 k_2\over (2\pi)^3}\ {{\rm d}^3 k_3\over (2\pi)^3} ,
\label{mamp0a}
\end{eqnarray}
where we dropped the explicit spin state dependencies and the summation over all the spin states, because we either consider type I or type II separately, so that $g$ can either be $g_{ooe}$ or $g_{eoe}$ as given in Appen.~\ref{pmc}. For type I phase matching  
\begin{equation}
\Delta k_z = n_{eff}(\theta_X,\omega_3,\theta_3,\phi_3) k_{z3} - n_o(\omega_1) k_{z1}  - n_o(\omega_2) k_{z2} ,
\label{dkzia}
\end{equation}
and for type II phase matching  
\begin{eqnarray}
\Delta k_z & = & n_{eff}(\theta_X,\omega_3,\theta_3,\phi_3) k_{z3} - n_{eff}(\theta_X,\omega_1,\theta_1,\phi_1) k_{z1} \nonumber \\ & & - n_o(\omega_2) k_{z2} .
\label{dkziia}
\end{eqnarray}

The expression in Eq.~(\ref{mamp0a}) is quite general, but it may not be convenient for calculations. Therefore, we subsequently introduce a number of simplifications.

\section{\label{simple}Simplifications}

\subsection{Choosing a propagation direction}

Due to the vacuum dispersion relation, any three of the four quantities $\omega$, $k_x$, $k_y$ and $k_z$ will fix the remaining quantity. Therefore, the phase space integrals for propagating (on shell) fields only run over three of these quantities, usually $k_x$, $k_y$ and $k_z$. For an optical beam, continuously propagating through a linear medium, it is more convenient to specify $\omega$, $k_x$ and $k_y$ and thereby fix $k_z$. The reason is that the system's evolution is considered as a function of the propagation distance and not as a function of time. The input is therefore specified for all time as a two-dimensional function on a plane perpendicular to the propagation direction. Such an input can also be expressed in the Fourier domain as a two-dimensional function of $k_x$ and $k_y$ for a particular $\omega$. (This implies a monochromatic approximation, however one can also specify the input as a function of $\omega$.) In other words, we prefer to integrate over $\omega$ instead of $k_z$. The orthogonal basis for the quantum states is then defined by $|{\bf K},\omega,s\rangle$ instead of $|{\bf k},s\rangle$, where 
\begin{eqnarray}
{\bf K} & = & k_x \hat{x} + k_y \hat{y}
\label{def2dk} \\
{\bf k} & = & k_x \hat{x} + k_y \hat{y} + k_z \hat{z}
\label{def3dk}
\end{eqnarray}
and $s$ denotes the spin state. By fixing a specific direction for propagation (the $z$-direction) we explicitly break rotation invariance and, by implication, also Lorentz invariance.

From the vacuum dispersion relation it follows that
\begin{equation}
{\rm d}k_z = {\omega\over c^2 k_z}\ {\rm d}\omega .
\label{wvskz}
\end{equation}
In the applications that we consider the angular spectrum of the beam only contain nonzero components in the positive $k_z$ region. Since both sides of the $\omega$ axis map into the positive side of the $k_z$-axis the change of integration variables in Eq.~(\ref{wvskz}) is well defined. Applying this change of variables to Eq.~(\ref{mamp0a}), we obtain
\begin{eqnarray}
{\cal M} & = & c^{3/2} L \int \Psi_s^*({\bf K}_1,\omega_1) \Psi_i^*({\bf K}_2,\omega_2) \Psi_p( {\bf K}_3,\omega_3) \nonumber \\ 
& & \times g({\bf k}_1,{\bf k}_2,{\bf k}_3) {\omega_1 \omega_2\over \omega_3} \nonumber \\ 
& & \times (2\pi)^3 \delta(\Delta \omega) \delta(\Delta k_x) \delta(\Delta k_y) S(\Delta k_z L) \nonumber \\ 
& & \times {{\rm d}^2 k_1~{\rm d} \omega_1\over c^2k_{z1}(2\pi)^3}\ {{\rm d}^2 k_2~{\rm d} \omega_2\over c^2k_{z2}(2\pi)^3}\ {{\rm d}^2 k_3~{\rm d} \omega_3\over c^2k_{z3}(2\pi)^3}
\label{mamp0}
\end{eqnarray}
where 
\begin{eqnarray}
k_{z1} & = & \sqrt{{\omega_1^2\over c^2} -k_{x1}^2 - k_{y1}^2} \nonumber \\
k_{z2} & = & \sqrt{{\omega_2^2\over c^2} -k_{x2}^2 - k_{y2}^2} \nonumber \\
k_{z3} & = & \sqrt{{\omega_3^2\over c^2} -k_{x3}^2 - k_{y3}^2} .
\label{kzs}
\end{eqnarray}
For type I phase matching  
\begin{eqnarray}
\Delta k_z & = & n_{eff}(\theta_X,\omega_3,\theta_3,\phi_3) \sqrt{{\omega_3^2\over c^2} - k_{x3}^2 - k_{y3}^2} \nonumber \\ 
& & - n_o(\omega_1) \sqrt{{\omega_1^2\over c^2} -k_{x1}^2 - k_{y1}^2} \nonumber \\ 
& & - n_o(\omega_2) \sqrt{{\omega_2^2\over c^2} -k_{x2}^2 - k_{y2}^2} ,
\label{dkzi}
\end{eqnarray}
and for type II phase matching  
\begin{eqnarray}
\Delta k_z & = & n_{eff}(\theta_X,\omega_3,\theta_3,\phi_3) \sqrt{{\omega_3^2\over c^2} - k_{x3}^2 - k_{y3}^2} \nonumber \\ 
& & - n_{eff}(\theta_X,\omega_1,\theta_1,\phi_1) \sqrt{{\omega_1^2\over c^2} -k_{x1}^2 - k_{y1}^2} \nonumber \\ 
& & - n_o(\omega_2) \sqrt{{\omega_2^2\over c^2} -k_{x2}^2 - k_{y2}^2} .
\label{dkzii}
\end{eqnarray}

\subsection{Monochromatic pump}

We now assume that the pump is monochromatic. For this purpose we define the momentum space wave function of the pump by
\begin{equation}
\psi_p({\bf K},\omega)= c\sqrt{k_z}\ G({\bf K}) H(\omega-\omega_p;\delta\omega) .
\label{pompdef}
\end{equation}
where $\omega_p$ and $\delta\omega$ are, respectively, the center frequency and the (small) bandwidth of the pump laser. (One cannot define $H(\omega)$ as a Dirac delta function, because that would lead to a squared Dirac delta function under the integral in the normalization condition, which gives a divergent result.) To satisfy the normalization requirement for the momentum space wave function
\begin{equation}
\int \left| \psi_p({\bf K},\omega) \right|^2\ {{\rm d}^2 k~{\rm d} \omega\over c^2k_z (2\pi)^3} = 1 ,
\label{norm0}
\end{equation}
we define
\begin{equation}
H(\omega;\delta\omega)= {2^{1/2} \pi^{1/4} \over \delta\omega^{1/2}} \exp\left[-{\omega^2\over 2\delta\omega^2}\right] .
\label{hdef}
\end{equation}
The frequency spectrum in Eq.~(\ref{hdef}) does not actually enforce the pump to be monochromatic, unless we take $\delta\omega\rightarrow 0$. 

The normalization now reduces to
\begin{equation}
\int \left| \psi_p({\bf K},\omega) \right|^2\ {{\rm d}^2 k~{\rm d} \omega\over c^2k_z (2\pi)^3} = 
\int \left| G({\bf K}) \right|^2 {{\rm d}^2 k\over (2\pi)^2} = 1 ,
\label{norm1}
\end{equation}
where we use the fact that
\begin{equation}
\int \left| H(\omega-\omega_p;\delta\omega) \right|^2 {{\rm d} \omega\over 2\pi}  = 2\pi .
\label{normh}
\end{equation}

Substituting Eq.~(\ref{pompdef}) into Eq.~(\ref{mamp0}) and evaluating the integral over $\omega_3$, we obtain
\begin{eqnarray}
{\cal M} & = & c^{3/2} L \int \Psi_s^*({\bf K}_1,\omega_1) \Psi_i^*({\bf K}_2,\omega_2) G({\bf K}_3) \nonumber \\ & & \times g({\bf k}_1,{\bf k}_2,{\bf k}_3) H(\omega_1+\omega_2-\omega_p;\delta\omega) {\omega_1 \omega_2 \over \omega_1+\omega_2} \nonumber \\ & & \times \delta(\Delta k_x) \delta(\Delta k_y) S(\Delta k_z L)  \nonumber \\ & & \times {{\rm d}^2 k_3\over c\sqrt{k_{z3}}}\ {{\rm d}^2 k_1~{\rm d} \omega_1\over c^2k_{z1}(2\pi)^3}\ {{\rm d}^2 k_2~{\rm d} \omega_2\over c^2k_{z2}(2\pi)^3}
\label{mamp1}
\end{eqnarray}
where $\Delta k_z$ is given by Eqs.~(\ref{dkzi}) and (\ref{dkzii}) with the replacement $\omega_3\rightarrow\omega_1+\omega_2$, and
\begin{equation}
k_{z3} = \sqrt{{(\omega_1+\omega_2)^2\over c^2} - k_{x3}^2 - k_{y3}^2} .
\label{kz3}
\end{equation}

\subsection{Degenerate, collinear type I phase matching}

At this point we restrict ourselves to type I phase matching, with degenerate signal and idler frequencies ($\omega_1=\omega_2=\omega_d=\omega_p/2$) and collinear signal and idler beams. Due to the phase matching condition for collinear beams, the effective refractive index for the pump beam must be equal to the ordinary index at the degenerate frequency [$n_{eff}(\theta_X,\omega_p,0,0) = n_o(\omega_d)$]. We'll therefore denote all refractive indices simply by $n_o$. Moreover, assuming that the nonlinear coefficient is a slow varying function of the three propagation vectors and that the beams are paraxial, we replace $g({\bf k}_1,{\bf k}_2,{\bf k}_3)$ with a constant $g$ and pull it out of the integral.

The degeneracy condition is implemented by replacing 
\begin{eqnarray}
\Psi_s^*({\bf K}_1,\omega_1) & = & c \sqrt{k_{z1}} M_s^*({\bf K}_1) H(\omega_1-\omega_d;\delta\omega_f) \nonumber \\
\Psi_i^*({\bf K}_2,\omega_2) & = & c \sqrt{k_{z2}} M_i^*({\bf K}_2) H(\omega_2-\omega_d;\delta\omega_f)
\label{redefsi}
\end{eqnarray}
in Eq.~(\ref{mamp1}), where $H(\cdot)$ is defined in Eq.~(\ref{hdef}) and $\delta\omega_f$ is the bandwidth of the line filter that imposes degeneracy. Assuming that the bandwidths are small enough one can simply substitute $\omega_1=\omega_2=\omega_d$ everywhere, except in the $H$-functions. Evaluating the integrals over $\omega_1$ and $\omega_2$, one obtains
\begin{widetext}
\begin{equation}
\int H(\omega_1-\omega_d;\delta\omega_f) H(\omega_2-\omega_d;\delta\omega_f) H(\omega_1+\omega_2-2\omega_d;\delta\omega)\ {{\rm d}\omega_1\over 2\pi}\ {{\rm d}\omega_2\over 2\pi} ={\sqrt{2}\over\pi^{1/4}} {\delta\omega_f \sqrt{\delta\omega}\over \sqrt{\delta\omega^2+2\delta\omega_f^2}} \approx {\sqrt{\delta\omega}\over\pi^{1/4}} ,
\label{haas}
\end{equation}
\end{widetext}
where the approximation follows from the fact that the line filter bandwidth is usually much larger than the pump bandwidth $\delta\omega \ll \delta\omega_f$.

In the paraxial limit the remaining $k_z$-factors become
\begin{equation}
\left(k_{z1} k_{z2} k_{z3}\right)^{-1/2} \approx 2\left({c\over\omega_p}\right)^{3/2} \left[ 1 + O\left( \theta^2 \right) \right] .
\label{park}
\end{equation}
where $\theta(=\lambda_p/\pi d_0)$ is the pump beam angle, with $d_0$ being the radius of the pump beam profile at its waist. Therefore, we use only the leading order term for this factor.

The degenerate, collinear type I phase matching condition implies that 
\begin{equation}
\delta(\Delta k_x) \delta(\Delta k_y) = {1\over n_o^2} \delta({\bf K}_3 - {\bf K}_1 - {\bf K}_2) .
\label{deltas}
\end{equation}
Using Eqs.~(\ref{redefsi}), (\ref{haas}), (\ref{park}) and (\ref{deltas}), and evaluating the phase space integrals for the pump to eliminate the remaining Dirac deltas in Eq.~(\ref{mamp1}), we obtain
\begin{eqnarray}
{\cal M} & = & {L g \over 2\pi^{1/4} n_o^2} \sqrt{\delta\omega\over \omega_p} \int M_s^*({\bf K}_1) M_i^*({\bf K}_2) \nonumber \\ & & \times G({\bf K}_1 + {\bf K}_2) S(\Delta k_z L)\ {{\rm d}^2 k_1\over (2\pi)^2}\ {{\rm d}^2 k_2\over (2\pi)^2}
\label{mamp2}
\end{eqnarray}
where $\Delta k_z = n_o (k_{z3}-k_{z1}-k_{z1})$ and
\begin{eqnarray}
k_{z1} & = & \sqrt{{\omega_d^2\over c^2} - k_{x1}^2 - k_{y1}^2} \nonumber \\
k_{z2} & = & \sqrt{{\omega_d^2\over c^2} - k_{x2}^2 - k_{y2}^2} \nonumber \\
k_{z3} & = & \sqrt{{4 \omega_d^2\over c^2} - (k_{x1}+k_{x2})^2 - (k_{y1}+k_{y2})^2} .
\label{kzsd}
\end{eqnarray}

In the paraxial limit $\Delta k_z$ simplifies to 
\begin{equation}
\Delta k_z = n_o c {(k_{x1}-k_{x2})^2 + (k_{y1}-k_{y2})^2 \over 4 \omega_d } .
\label{dkzidc5}
\end{equation}
It is often more convenient to do the calculation in terms of spatial frequencies instead of the propagation vector components. Therefore, we define the propagation vector in terms of spatial frequencies by
\begin{equation}
{\bf K}=2\pi(a\hat{x}+b\hat{y}) .
\label{kab}
\end{equation}
Then $\Delta k_z$ becomes
\begin{equation}
\Delta k_z = {2 \kappa \over L} \left[ (a_1-a_2)^2 + (b_1-b_2)^2 \right] ,
\label{dkzidc6}
\end{equation}
where 
\begin{equation}
\kappa=\pi n_o\lambda_p L .
\label{kappa}
\end{equation}
Hence
\begin{equation}
S(\Delta k_z L) = {\sin\left\{ 2 \kappa \left[ (a_1-a_2)^2 + (b_1-b_2)^2 \right]\right\} \over 2 \kappa \left[ (a_1-a_2)^2 + (b_1-b_2)^2 \right]} .
\label{sfunc0}
\end{equation}

\section{\label{bgspek}Bessel-Gauss OAM spectrum}

We now evaluate the integrals in Eq.~(\ref{mamp2}) for a particular choice of momentum space wave functions for the pump, signal and idler beams, respectively. 
It is assumed that the pump beam has a Gaussian profile, so that
\begin{equation}
G({\bf K}) = \sqrt{2\pi} d_0 \exp\left[ -d_0^2\pi^2 \left( a^2+b^2 \right) \right] ,
\label{pomprof}
\end{equation} 
where $d_0$ is the beam waist radius. The momentum space wave function in Eq.~(\ref{pomprof}) satisfies the normalization condition given in Eq.~(\ref{norm1}).

The signal and idler beams are assumed to be BG modes. For convenience we'll use the generating function for the angular spectra of the BG modes (see Appen.~\ref{besgauss}), which is given by
\begin{eqnarray}
\tilde{\cal G}(a,b) & = & \sqrt{2\pi} d_n \exp \left\{ -\pi^2 d_n^2 (a^2+b^2) - d_n^2 h_n^2 /4 \right. \nonumber \\ 
& & \left. + \pi h_n d_n^2 [\cos(\alpha_n) b - \sin(\alpha_n) a] \right\} ,
\label{fgbg0}
\end{eqnarray}
where $a$ and $b$ are the spatial frequency coordinates, $h_n$ is the radial mode parameter of the BG modes, $d_n$ is the radius of the Gaussian envelope and $\alpha_n$ is the generating parameter which is used, according to Eq.~(\ref{gbgx}), to generate the angular spectrum of a BG mode with a particular azimuthal index $\ell$. The subscript $n(=1,2)$ represents the signal or idler beam. We set $z=0$ inside the generating function, thereby assuming that the waists of the BG modes are located in the nonlinear crystal where these modes are generated.

Substituting
\begin{equation}
M_s^*({\bf K}_1) M_i^*({\bf K}_2) = {\cal G}_s^*(a_1,b_1) {\cal G}_i^*(a_2,b_2) ,
\end{equation}
and Eqs.~(\ref{sfunc0}) and (\ref{pomprof}) into Eq.~(\ref{mamp2}), and using Eq.~(\ref{kab}), we obtain the integral expression for the generating function of the scattering amplitudes
\begin{eqnarray}
{\cal A} & = & \Omega_0 \int \exp\left\{ -d_0^2\pi^2 \left[ (a_1+a_2)^2+(b_1+b_2)^2 \right] \right\} \nonumber \\
& & \times {\sin\left\{ 2 \kappa \left[ (a_1-a_2)^2 + (b_1-b_2)^2 \right]\right\} \over 2 \kappa \left[ (a_1-a_2)^2 + (b_1-b_2)^2 \right]} \nonumber \\
& & \times {\cal G}_s^*(a_1,b_1) {\cal G}_i^*(a_2,b_2)\  {\rm d} a_1\ {\rm d} b_1\ {\rm d} a_2\ {\rm d} b_2 ,
\label{mamp3}
\end{eqnarray}
where
\begin{equation}
\Omega_0 = {\pi^{1/4} d_0 L g \over \sqrt{2} n_o^2} \sqrt{\delta\omega \over\omega_p} .
\label{voor0}
\end{equation}

\subsection{Result after integration}

To expedite the evaluation of the integrals in Eq.~(\ref{mamp3}), we'll assume that $h_1=h_2=h$, and also that $d_1=d_2=\omega_0$. Moreover, we assume that the pump wavelength $\lambda_p$ is much smaller than any of the other dimension parameters. Under these conditions the integrals evaluate to
\begin{eqnarray}
{\cal A} & = & {\Omega_0 \gamma^2 \over \pi d_0^2\eta(1+2\gamma^2)\xi^2\sigma^2} \exp\left[-{(\gamma^2+\sigma)\xi\over 1+2\gamma^2} \right] \nonumber \\
& & \times \left \{ i [1-(1+i\eta)\xi\sigma] \exp[(1+i\eta)\xi\sigma] \right. \nonumber \\ 
& & \left. - i [1-(1-i\eta)\xi\sigma]  \exp[(1-i\eta)\xi\sigma] \right \}
\label{mamp4}
\end{eqnarray}
where
\begin{eqnarray}
\gamma & = & {d_0\over\omega_0} \label{gamma} \\
\xi & = & h^2 \omega_0^2 \label{xi} \\
\eta & = & {n_o \lambda_p L\over\pi \omega_0^2} \label{eta} \\
\sigma & = & {1\over 2} \sin \left({\alpha_1-\alpha_2\over 2}\right)^2  \label{chi}
\end{eqnarray}
and
\begin{equation}
\Omega_1 = {L g \over \sqrt{2} \pi^{3/4} d_0 n_o^2} \sqrt{\delta\omega \over \omega_p} .
\label{voor1}
\end{equation}

The overall absolute magnitude of the probability for producing down-converted photon pairs is represented by the dimensionless quantity $\Omega_1$ given in Eq.~(\ref{voor1}). Apart from some numerical constants and the nonlinear coefficient $g$, it depends on the ratio of the length of the nonlinear crystal $L$ and the radius of the pump beam $d_0$, as well as on the square-root of the fraction of the bandwidth of the pump compared to the center frequency of the pump.

\subsection{Extraction of the different orders}

The coefficients for the different orders are extracted with the aid of Eq.~(\ref{gbgx}). Since the extraction needs to be done for both the signal and idler beams, the relevant expression is
\begin{eqnarray}
{\cal M}_{nm} & = & {1\over 4\pi^2} \int_0^{2\pi} \!\! \int_0^{2\pi} \exp(in\alpha_1) \exp(im\alpha_2) \nonumber \\
& & \times {\cal A}(\alpha_1-\alpha_2)\ {\rm d}\alpha_1\ {\rm d}\alpha_2 .
\label{gbgxx}
\end{eqnarray}
However, since ${\cal A}$ only depends of the difference in angle $\alpha_1-\alpha_2$ and since these angles are cyclic variables, one can redefine one of the angles as $\alpha_2=\alpha_1-\alpha_3$, so that ${\cal A}$ becomes independent of $\alpha_1$. The integral over $\alpha_1$ then gives a Kronecker delta function, 
\begin{equation}
{1\over 2\pi}\int_0^{2\pi} \exp[-i(n+m)\alpha_1]\ {\rm d}\alpha_1 = \delta_{-n,m} ,
\label{conserv}
\end{equation}
which shows that orbital angular momentum is conserved in the SPDC process. The remaining integral over $\alpha_3$ is given by
\begin{equation}
{\cal M}_m = {1\over 2\pi}\int_0^{2\pi} {\cal A}(\alpha_3) \exp(-im\alpha_3)\ {\rm d}\alpha_3 .
\label{gbgx0}
\end{equation}

\begin{figure}[ht]
\centerline{\scalebox{1}{\includegraphics{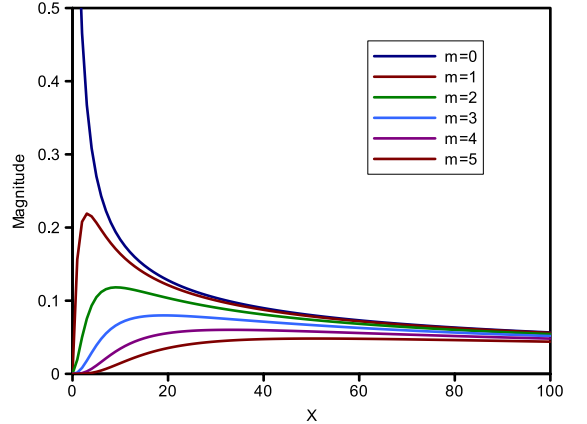}}}
\caption{The curves for the normalized magnitudes of the coefficients $|{\cal M}_m(X)|$ as a function of $X$ for a number of different coefficients ($m=0...5$).}
\label{xgraf}
\end{figure}

Upon evaluating the integral over $\alpha_3$, we obtain
\begin{widetext}
\begin{eqnarray}
{\cal M}_m & = & (-1)^m \Omega_1 \exp\left(-\frac{X}{2}\right) \left \{ {2\gamma^2\over 1+2\gamma^2} I_m\left(\frac{X}{2}\right) - {\eta^2\over 24\gamma^2} X [X(1+2\gamma^2)+2\gamma^2-1] I_{m+1}\left(\frac{X}{2}\right) \right. \nonumber \\ 
& & \left. - {\eta^2\over 24\gamma^2} [(1+2\gamma^2) m^2+ (1+2\gamma^2) (m+X)^2 + 4X\gamma^2 + 2(2\gamma^2-1) m ] I_m\left(\frac{X}{2}\right)   \right\} ,
\label{koef0}
\end{eqnarray}
\end{widetext}
up to second order in $\eta$, where $I_m(\cdot)$ is the modified Bessel function and 
\begin{equation}
X = {\xi\gamma^2 \over 1+2\gamma^2} = {h^2 w_0^2 d_0^2 \over w_0^2+2 d_0^2} .
\label{xdef}
\end{equation}

Ignoring the $\eta^2$-terms in Eq.~(\ref{koef0}), we see that the dominent behavior of the coefficients is given by
\begin{equation}
\left| {\cal M}_m (X) \right|  \sim  \exp\left(-\frac{X}{2}\right) I_m\left(\frac{X}{2}\right) .
\label{msim}
\end{equation}

\begin{figure}[ht]
\centerline{\scalebox{1}{\includegraphics{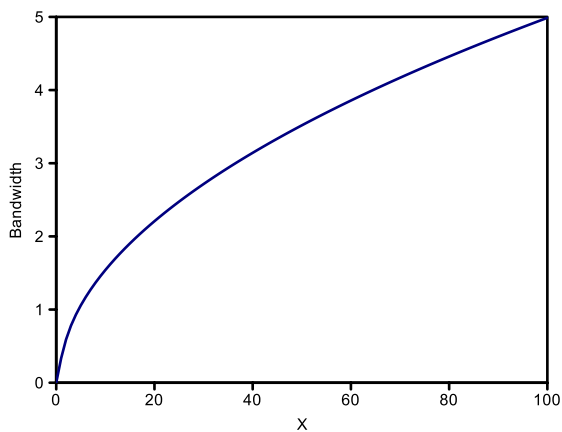}}}
\caption{The curves for the OAM bandwidth as defined in Eq.~(\ref{bwdef}) as a function of $X$.}
\label{bandwfig}
\end{figure}

In Fig.~\ref{xgraf} we show the curves for Eq.~(\ref{msim}) as a function of $X$ for a number of different coefficients. For $X=0$ the first coefficient ($m=0$) is 1 and the rest are zero. As $X$ becomes larger the magnitudes of the coefficients become closer to each other. It is therefore desirable to make $X$ as large as possible.

\subsection{OAM Bandwidth}

One can compute the OAM bandwidth of the down-converted beams by exploiting the generating function in Eq.~(\ref{mamp4}). We define the OAM bandwidth by 
\begin{equation}
B = \left( {\sum m^2 \left| {\cal M}_m \right|^2 \over \sum \left| {\cal M}_m \right|^2} \right)^{1/2} .
\label{bwdef}
\end{equation}
Formally the generating function for the coefficients can be written as
\begin{equation}
{\cal A}(\alpha) = \sum_{m=-\infty}^{\infty} \exp(im\alpha) {\cal M}_m .
\label{genkoef}
\end{equation}
The conservation of angular momentum of Eq.~(\ref{conserv}) turns the double summation into a single summation.

The bandwidth as defined in Eq.~(\ref{bwdef}) can be computed directly from the generating function in Eq.~(\ref{mamp4}) as follows
\begin{equation}
B = \left( {\int \left| \partial_{\alpha} {\cal A}(\alpha) \right|^2\ {\rm d}\alpha \over \int \left| {\cal A}(\alpha) \right|^2\ {\rm d}\alpha} \right)^{1/2} .
\label{bandw}
\end{equation}

The resulting expression for the bandwidth, up to second order in $\eta$, is
\begin{widetext}
\begin{eqnarray}
B & = & \frac{1}{2} \left[ {X I_1(X)\over I_0(X)} \right]^{1/2} \left\{ 1 + {\eta^2(1+2\gamma^2)\over 96\gamma^4} \left[ 2(1+2\gamma^2) 
+ (4\gamma^2 X+2 X+6\gamma^2-1) {X I_1(X)\over I_0(X)} \right. \right. \nonumber \\
& & \left. \left. - (4\gamma^2 X+2 X+10\gamma^2+1) {X I_0(X)\over I_1(X)} \right] \right\}
\label{bandw0}
\end{eqnarray}
\end{widetext}
where $X$ is given in Eq.~(\ref{xdef}). The leading order term gives the bandwidth purely as a function of $X$. In Fig.~\ref{bandwfig} we show the curves for the OAM bandwidth as defined in Eq.~(\ref{bwdef}) as a function of $X$. We see that the bandwidth is a monotonically increasing function of $X$, which confirms the observation at the end of the previous section. Since $X$ is proportional to the square of the transverse scale parameter for the Bessel functions $h$, we find that a large beandwidth is obtained for large values of $h$.

\section{\label{concl}Conclusions}

The OAM spectrum for BG modes generated by the SPDC process is calculated. This calculation is preceeded by a general derivation of expressions for the spectrum generated by the SPDC process in any transverse spatial modal basis. Through a series of simplifications, we eventually arrive at expressions for the OAM spectrum for BG modes in degenerate, collinear SPDC with type I phase matching. The calculation of the OAM spectrum is done using generating functions for the angular spectra of the BG modes. We obtained a closed form expression for the bandwidth of this OAM spectrum. The results show that the magnitudes of the coefficients in the BG modal expansion of the quantum states of the down-converted photon in terms of become close to being equal, provided that the transverse scale parameter for the Bessel functions is as large as possible. 

The expressions for the OAM spectrum also provide the overall absolute magnitude of the spectrum, which is represented by the dimensionless quantity $\Omega_1$ in Eq.~(\ref{voor1}), however we did not investigate this quantity any further.

The next step would be to perform physical experiments in which the coefficients for the different sets of modes are measured. In this way one can check whether the shape and the absolute magnitude of the observed spectra agree with the theoretical results of this paper.

\acknowledgments

The author wishes to extends his gratitude to Robert Boyd, Andrew Forbes and Miles Padgett for fruitful discussions on this topic. The work has been done with the support of a SRP Type A grant from the CSIR.

\appendix

\section{\label{pmc}Effective nonlinear coefficients}

Considering the case of a negative uniaxial crystal for which extraordinary index $n_e$ is smaller than the ordinary index $n_o$, one obtains the following two phase matching conditions:
\begin{itemize}
\item Type I phase matching, for which
\begin{eqnarray}
g_{ooe} & = & \sqrt{n_{eff}(\theta_X,\omega_3,\theta_3,\phi_3) n_o(\omega_1) n_o(\omega_2)} \nonumber \\ & & \times 3 \omega_3^2 \sqrt{\hbar\over 2 c^3 \epsilon_0} \chi^{(2)}_{abc} \eta_o^{a*}(\hat{k}_1) \eta_o^{b*}(\hat{k}_2) \eta_e^c(\hat{k}_3)
\label{gooe}
\end{eqnarray}
and
\begin{eqnarray}
\Delta {\bf k}_I & = & n_{eff}(\theta_X,\omega_3,\theta_3,\phi_3){\bf k}_3 \nonumber \\
 & & - n_o(\omega_1){\bf k}_1 - n_o(\omega_2){\bf k}_2
\end{eqnarray}
\item Type II phase matching, for which
\begin{eqnarray}
g_{eoe} & = & \sqrt{n_{eff}(\theta_X,\omega_3,\theta_3,\phi_3) n_{eff}(\theta_X,\omega_1,\theta_1,\phi_1) n_o(\omega_2)} \nonumber \\ & & \times 3 \omega_3^2 \sqrt{\hbar\over 2 c^3 \epsilon_0} \chi^{(2)}_{abc} \eta_e^{a*}(\hat{k}_1) \eta_o^{b*}(\hat{k}_2) \eta_e^c(\hat{k}_3)
\end{eqnarray}
and 
\begin{eqnarray}
\Delta {\bf k}_{II} & = & n_{eff}(\theta_X,\omega_3,\theta_3,\phi_3){\bf k}_3 \nonumber \\ 
& & - n_{eff}(\theta_X,\omega_1,\theta_1,\phi_1){\bf k}_1 - n_o(\omega_2){\bf k}_2 .
\label{dkii}
\end{eqnarray}
\end{itemize}
In the above expressions we define the propagation vectors by
\begin{equation}
{\bf k}=k\ (\hat{x}\sin\theta\cos\phi+\hat{y}\sin\theta\sin\phi+\hat{z}\cos\theta)
\end{equation}
and the optic axis by
\begin{equation}
\hat{a} = \hat{y}\sin\theta_X+\hat{z}\cos\theta_X .
\end{equation}
The effective refractive index $n_{eff}(\cdot)$ is given as 
\begin{widetext}
\begin{equation}
n_{eff}(\theta_X,\omega,\theta,\phi) = {n_e(\omega)  n_o(\omega) \over \sqrt{n_o^2(\omega) + [n_e^2(\omega)-n_o^2(\omega)](\sin\theta_X \sin\theta \sin\phi + \cos\theta_X \cos\theta)^2}} .
\label{vgneff}
\end{equation}
\end{widetext}
The parameters with the subscripts $1$, $2$ and $3$ in Eqs.~(\ref{gooe})-(\ref{dkii}) represent the parameters for the signal, idler and pump beams, respectively.

\section{\label{besgauss}Generating function for the Bessel-Gaussian modes}

In normalized coordinates the Bessel-Gaussian (BG) modes are given by 
\begin{eqnarray}
M^{\rm BG}_{\ell}(r,\phi,t;\chi) & = & \sqrt{2\over\pi} {\rm J}_{\ell} \left( {\chi r\over 1-it} \right) \exp(i\ell\phi) \nonumber \\
& & \times  \exp \left( {i \chi^2 t - 4 r^2 \over 4(1-it)} \right)
\label{bgm}
\end{eqnarray}
where the normalized radial coordinate is given by $r=(x^2+y^2)^{1/2}/\omega_0$, $\phi$ is the azimuthal coordinate and the normalized propagation distance is $t=z/z_R$; $\ell$ is the mode index (a signed integer) and $\chi(=h\omega_0)$ is a normalized parameter for the size of the Bessel mode. The initial radius of the Gaussian profile is $\omega_0$ and the Rayleigh range is $z_R=\pi\omega_0^2/\lambda$. There is an additional propagation phase factor associated with every BG mode given by $\exp(-ib z)$, which is not shown in Eq.~(\ref{bgm}). The wavenumber $k$ and the parameters $b$ and $h$ are related by $k^2 = b^2 + h^2$. So $b$ and $h$ are, respectively, the magnitudes of the longitudinal and transverse components of the dominant wave vectors in the spectrum of the BG mode.

The generating function for the BG modes for a particular $\chi$ is given by 
\begin{equation}
{\cal G} = \sqrt{2\over\pi} {1 \over 1-it} \exp \left( {i 4 \chi r \sin(\phi-\alpha) -4 r^2 + i \chi^2 t \over 4(1-it)} \right) ,
\label{gbg}
\end{equation}
where $\alpha$ is an angular generating parameter that is used to generate a BG mode with a specific value of the azimuthal index $\ell$. To generate such a mode one performs the following calculation
\begin{equation}
M^{\rm BG}_{\ell} = {1\over 2\pi} \int_0^{2\pi} {\cal G} \exp(i\ell\alpha)\ {\rm d}\alpha .
\label{gbgx}
\end{equation}

The Fourier transform of the generating function gives the generating function for the angular spectra of the BG modes and is given by 
\begin{eqnarray}
\tilde{\cal G} & = & {\cal F} \{ {\cal G} \} = \int_{-\infty}^{\infty} {\cal G} \exp[i2\pi(a u + b v)]\ {\rm d} a\ {\rm d} b \nonumber \\
& = & \sqrt{2\pi} \exp \left[ -\pi^2 \mu^2 (1-it) + \pi \chi \mu \sin(\nu-\alpha) - {\chi^2\over 4} \right] , \nonumber \\
\label{fgbg}
\end{eqnarray}
where $\mu$ and $\nu$ are, respectively, the normalized radial coordinate and the angular coordinate for the spatial frequency domain. The angular spectra of the BG modes are generated from Eq.~(\ref{fgbg}) in the same way as in Eq.~(\ref{gbgx}).

%\bibliography{bessprl}

\end{document}